\documentclass{article}[12pt]
\usepackage{amsmath,amssymb}
\usepackage{graphicx}%
\usepackage{verbatim}

\begin{document}

\begin{center}
{\bf \Large Qualitative and Numerical Analysis of \\[6pt]
a Cosmological Model Based on an Asymmetric\\[6pt]
Scalar Doublet with Minimal connections.\\[6pt]
III. Multiply-connected Factor\\[6pt] and Character of the Singular Points        } \\[12pt]
Yu. G. Ignat'ev and I. A. Kokh\\
N. I. Lobachevsky Institute of Mathematics and Mechanics of Kazan Federal University,\\
Kremleovskaya str., 35, Kazan, 420008, Russia.
\end{center}


\begin{abstract}

On the basis of a qualitative and numerical analysis of a cosmological model based on an asymmetric scalar doublet of nonlinear, minimally interacting scalar fields, both classical and phantom, the behavior of the model near zero energy hypersurfaces has been revealed.  The influence of the multiply connected factor of the phase space of the dynamical system, this factor being a consequence of the nonanalyticity of the coefficients of an autonomous system of differential equations, is discussed.  The character of all singular points is revealed.\\


\end{abstract}

\section{Introduction}

\noindent In preceding papers, we performed a qualitative analysis of a cosmological model based on an asymmetric doublet [1] and also carried out analytical and numerical investigations of its behavior near  hypersurfaces of zero effective energy [2], revealing a tendency of the phase trajectories to wind onto this hypersurface in the presence of attractive foci inside it.  In particular, we showed that for appropriate values of the model parameters, motion on the zero effective energy surface is an exact limit solution of the dynamical equations of the model.  The dynamical system investigated here is described by a normal autonomous system of ordinary differential equations [1]:
\begin{equation} \label{_1_}
\begin{array}{c} {\Phi '=Z,} \\[12pt] {Z'=-\sqrt{3} Z\sqrt{\left(Z^{2} +e\Phi ^{2} -{\displaystyle\frac{\alpha _{m} }{2}} \Phi ^{4} \right)-\left(z^{2} -\varepsilon \mu ^{2} \varphi ^{2} +{\displaystyle\frac{\beta _{m} }{2}} \varphi ^{4} \right)+\lambda _{m} } -e\Phi +\alpha _{m} \Phi ^{3} ,} \\[12pt] {\varphi '=z,} \\[12pt] {z'=-\sqrt{3} z\sqrt{\left(Z^{2} +e\Phi ^{2} -{\displaystyle\frac{\alpha _{m} }{2}} \Phi ^{4} \right)-\left(z^{2} -\varepsilon \mu ^{2} \varphi ^{2} +{\displaystyle\frac{\beta _{m} }{2}} \varphi ^{4} \right)+\lambda _{m} } +\varepsilon \mu ^{2} \varphi -\beta _{m} \varphi ^{3} ,} \end{array}
\end{equation}
where
\[\alpha _{m} =\frac{\alpha }{m^{2} } ,\quad \beta _{m} =\frac{\beta }{m^{2} } ,\quad \lambda _{m} =\frac{\lambda }{m^{2} } ,\quad \Lambda _{m} =\lambda _{m} -\frac{1}{2\alpha _{m} } -\frac{\mu ^{2} }{2\beta _{m} } ,\quad \mu \equiv \frac{{\rm {\tt m}}}{m} \]
are the normalized dimensionless parameters of the model, and which we will refer to in the form of a list {\bf P}:
\[{\bf P}\equiv [\alpha _{m} ,\beta _{m} ,e,\varepsilon ,\mu ,\lambda _{m} ],\quad {\bf I}\equiv [\Phi (0),Z(0),\varphi (0),z(0)],\]
where {\bf I} is a list of initial conditions${}^{\ }$\footnote{\ Since\ system\ (1)\ is\ invariant\ with\ respect\ to\ translations\ of\ dimensionless\ time $\tau$,\ the\ choice\ of\ the\ initial\ time\ $ \tau_0=0$ is\ irrelevant.\ }.

In this paper, we perform a detailed study of the cosmological evolution of an asymmetric\footnote{\ That\ is,\ consisting\ of\ a\ pair\ of\ scalar\ fields\ of\ different\ nature:\ classical\ and\ phantom.\ \ \ \ \ } scalar doublet as a function of the parameters of the model.

It is necessary to augment system \eqref{_1_} by the Einstein equation, which determines the dimensionless Hubble constant $H_{m}$:
\begin{equation} \label{_2_}
\frac{a'^{2} }{a^{2} } \equiv H_{m}^{2} =\frac{1}{3} {\rm {\mathcal E}}_{m} (\Phi ,Z,\varphi ,z),
\end{equation}
and also to define the invariant cosmological acceleration $\Omega$:
\begin{equation} \label{_3_}
H(t)=\frac{\dot{a}}{a} \ge 0;\; \; \Omega (t)=\frac{a\ddot{a}}{\dot{a}^{2} } \equiv 1+\frac{\dot{H}}{H^{2} } =-\frac{1}{2} (1+3{\rm \varkappa }),
\end{equation}
where $\varkappa=p_m/ {\mathcal E}_m$ is the ratio of the effective pressure to the effective energy density,\footnote{\ With\ the\ cosmological\ constant\ taken\ into\ account\ (see\ [4]).\ \ } this ratio being known as the effective barotropic coefficient.  The effective energy density and pressure of the dynamical system have the following form:
\[\begin{array}{l} {{\rm {\mathcal E}}_{m} (\Phi ,Z,\varphi ,z)\equiv \left(Z^{2} +e\Phi ^{2} -{\displaystyle\frac{\alpha _{m} }{2} }\Phi ^{4} \right)+\left(-z^{2} +\varepsilon \mu ^{2} \varphi ^{2} -{\displaystyle\frac{\beta _{m} }{2}} \varphi ^{4} \right)+\lambda _{m} ,} \\ {{\rm }p_{m} (\Phi ,Z,\varphi ,z)\equiv \left(Z^{2} -e\Phi ^{2} +{\displaystyle\frac{\alpha _{m} }{2}} \Phi ^{4} \right)-\left(z^{2} +\varepsilon \mu ^{2} \varphi ^{2} -{\displaystyle\frac{\beta _{m} }{2}} \varphi ^{4} \right)-\lambda _{m} .} \end{array}\]

\section{ Influence of multiple connection of the phase space on the behavior of the dynamical system   }

\noindent In order for system of differential equations \eqref{_1_} to have a real solution, it is necessary that the radicand in the equations be nonnegative, i.e., that the effective energy of the system with the cosmological constant taken into account be nonnegative:
\begin{equation} \label{_4_}
{\rm {\mathcal E}}_{m} (\Phi ,Z,\varphi ,z)\equiv \left(Z^{2} +e\Phi ^{2} -\frac{\alpha _{m} }{2} \Phi ^{4} \right)+\left(-z^{2} +\varepsilon \mu ^{2} \varphi ^{2} -\frac{\beta _{m} }{2} \varphi ^{4} \right)+\lambda _{m} \ge 0.
\end{equation}
Inequality \eqref{_4_} can lead to violation of single connection of the phase space and formation in it of closed lacunae, bounded by surfaces with zero effective energy.  It can be expected that in the case in which attractive centers are found inside forbidden regions, the phase trajectories will wind onto the zero effective energy surfaces, whereas in the case in which saddle points are found in these regions, the phase trajectories will be repulsed from a zero effective energy surface. Note, however, another circumstance qualitatively distinguishing a cosmological model with an asymmetric scalar doublet from the corresponding model with a single scalar field, considered above. The hypersurface of phase space  $S_3^0\in \mathbb{R}_4$ is determined only by the parameters of the field model of the scalar doublet $\{\alpha _{m} ,\beta _{m} ,e,\varepsilon ,\mu ,\lambda _{m}\}$ and does not depend on the time variable $\tau$. However, intersections of two-dimensional phase surfaces of the solitary fields $\Sigma_\Phi=\{\Phi,Z\}$ and $\Sigma_\varphi=\{\varphi,z\}$ with the hypersurface $S_3^0$ can be two-dimensional curves $\Gamma_\Phi$ and $\Gamma_\varphi$(closed and open), depending substantially on the values of the dynamical variables of the other scalar field, and therefore also depending on the time variables:
\begin{equation} \label{_5_}
\begin{array}{c}\displaystyle {\Gamma _{\Phi } (\tau ):\; {\rm {\mathcal E}}_{m} (\Phi ,Z,\tau )=0\Rightarrow {\rm {\mathcal E}}_{m} (\Phi ,Z,\varphi (\tau ),z(\tau ))=0,} \\[12pt] \displaystyle {\Gamma _{\varphi } (\tau ):\; {\rm {\mathcal E}}_{m} (\varphi ,z,\tau )=0\Rightarrow {\rm {\mathcal E}}_{m} (\Phi (\tau ),Z(\tau ),\varphi ,z)=0.} \end{array}
\end{equation}
As a consequence of Eqs. \eqref{_5_}, the topology of the two-dimensional phase subspaces  $\Sigma_\Phi$ and $\Sigma_\varphi$ can vary significantly in time.  This factor is fundamentally new and important for the cosmological model.

\section{ Singular points of the dynamical system     }

\noindent In [1] all the singular points of dynamical system \eqref{_1_} were determined and their character was indicated in general terms.  Let us briefly recapitulate all the singular points and indicate their character with allowance for the factor of their possible inaccessibility. We note the following important circumstance.  In the case in which the realness condition stated by Eq. \eqref{_4_} is fulfilled at a specific singular point $M_i$, phase trajectories can, in principle, arrive at such a singular point or depart from it.  If the singular point is found in a forbidden region of phase space, the phase trajectories of the dynamical system cannot pass through this singular point, but only be attracted to the boundary of the forbidden region or be repulsed from it, depending on the character of the singular point.

\noindent Thus, as we indicated earlier, dynamical system \eqref{_1_} has nine singular points:

\noindent 1)\textbf{ }\textit{M}${}_{0}$: For any values $\alpha_m$ and $\beta_m$ of  there is always a central singular point:
\begin{equation} \label{_6_}
x=0,\quad Z=0,\quad y=0,\quad z=0\Rightarrow M_{0} :(0,0,0,0).
\end{equation}
Substituting the obtained solution given by Eqs. \eqref{_6_} into condition \eqref{_4_}, we obtain the necessary condition for realness of the solutions at the singular point:
\[\lambda _{m} \ge 0.  \]
2) \textit{M}${}_{01}$, \textit{M}${}_{02}$: For any values of  $\alpha_m$ and $\varepsilon\beta_m>0$ we have two more points, symmetric in $\varphi$:
\begin{equation} \label{_7_}
x=0,\quad Z=0,\quad y_{\pm } =\pm \frac{\mu }{\sqrt{\varepsilon \beta _{m} } } ,\quad z=0\quad \Rightarrow M_{01} (0,0,|y_{\pm } |,0),\quad M_{02} (0,0,-|y_{\pm } |,0).
\end{equation}
A necessary condition of realness of the solutions at the singular points \textit{M}${}_{01}$ and \textit{M}${}_{02}$ is \textbf{\footnote{\ Here\ and\ below,\ we\ establish\ a\ correspondence\ between\ the\ notation\ of\ the\ parameters\ adopted\ in\ [1,\ 2]\   $u,v,w$  \ and\ the\ notation\ of\ the\ corresponding\ parameters\ in\ [3,\ 4]\   $\sigma _{1}^{2} ,\sigma _{2}^{2} ,\sigma _{3}^{2} $  .\ }}
\begin{equation} \label{_8_}
\sigma _{1}^{2} {\rm \; }(=u)\equiv \frac{\mu ^{4} }{2\beta _{m} } +\lambda _{m} \ge 0.
\end{equation}
3)\textbf{ }\textit{M}${}_{10}$, \textit{M}${}_{20}$: For any values of  $\beta_m$ and $e\alpha_m>0$ we have two more points symmetric in $\Phi$:
\begin{equation} \label{_9_}
x_{\pm } =\pm \frac{1}{\sqrt{e\alpha _{m} } } ,\quad Z=0,\quad y=0,\quad z=0\quad \Rightarrow M_{10} (|x_{\pm } |,0,0,0),\quad M_{20} (-|x_{\pm } |,0,0,0).
\end{equation}
A necessary condition for realness of the solutions at the singular points \textit{M}${}_{10}$ and \textit{M}${}_{20}$ is
\begin{equation} \label{_10_}
\sigma _{2}^{2} {\rm \; }\left(=\frac{3}{4} v\right)\equiv \frac{3}{4} \left(\frac{1}{2\alpha _{m} } +\lambda _{m} \right)\ge 0.
\end{equation}
4)\textbf{ }\textit{M}${}_{12}$, \textit{M}${}_{21}$, \textit{M}${}_{11}$, \textit{M}${}_{22}$:  For  $e\alpha_m>0$ and   $\varepsilon\beta_m>0$ we have four more points symmetric in $\Phi$  and $\varphi$:
\begin{equation} \label{_11_}
\begin{array}{c} {x_{\pm } =\pm {\displaystyle\frac{1}{\sqrt{e\alpha _{m} } }},\quad Z=0,\quad y_{\pm } =\pm {\displaystyle\frac{\mu }{\sqrt{\varepsilon \beta _{m} }} } ,\quad z=0\quad} \Rightarrow  \\[12pt] {M_{11} (|x_{\pm } |,0,|y_{\pm } |,0),\quad M_{12} (|x_{\pm } |,0,-|y_{\pm } |,0),\quad M_{21} (-|x_{\pm } |,0,|y_{\pm } |,0),\quad M_{22} (-|x_{\pm } |,0,-|y_{\pm } |,0).} \end{array}
\end{equation}
A necessary condition for realness of the solutions at the singular points \textit{M}${}_{12}$, \textit{M}${}_{21}$, \textit{M}${}_{11}$, and \textit{M}${}_{22}$ is
\begin{equation} \label{_12_}
\sigma _{3}^{2} \left(=\frac{3}{4} w\right)\equiv \frac{3}{4} \left(\frac{1}{2\alpha _{m} } +\frac{\mu ^{4} }{2\beta _{m} } +\lambda _{m} \right)\ge 0.
\end{equation}

\section{Character of the singular points of the dynamical system of an asymmetric scalar doublet     }

The minimal character of the interaction of the components of the doublet uniquely leads to a block-diagonal structure of the matrix of dynamical system \eqref{_1_}, which for $Z=z=0$ has the form
\begin{equation} \label{_13_}
A_{M} ={\displaystyle\left\| \frac{\partial F_{i} }{\partial x_{k} } \right\| _{M} =\left(\begin{array}{cccc} {0} & {1} & {0} & {0} \\ {\displaystyle{\frac{\partial Q}{\partial x}} } & {\displaystyle{\frac{\partial Q}{\partial Z}} } & {0} & {0} \\ {0} & {0} & {0} & {1} \\ {0} & {0} &{\displaystyle {\frac{\partial q}{\partial y} }} & {\displaystyle{\frac{\partial q}{\partial z} }} \end{array}\right)_{M}} .
\end{equation}
The determinant of this block-diagonal matrix is equal to
\begin{equation} \label{_14_}
\Delta (A)=\frac{\partial Q}{\partial x} \frac{\partial q}{\partial y} .
\end{equation}

Note that since all the dynamical variables are real quantities, the functions $Q$ and $q$ together with their partial derivatives in the forbidden regions of phase space are also real quantities.  However, in the forbidden regions of phase space, i.e., in the regions with negative effective energy ${\rm {\mathcal E}}_{m}<0$, the derivatives with respect to the dynamical variables can become imaginary quantities.  This means that the given singular point is found in an inaccessible region of phase space.  Note that in the matrix of the dynamical system, the realness condition can be violated, and simultaneously in derivatives of the type  ${\partial Q}/{\partial Z}$, ${\partial Q}/{\partial z}$, ${\partial q}/{\partial Z}$, ${\partial q}/{\partial z}$.

Let us consider the equations  for the eigenvectors ${\bf u}_i$ and eigenvalues $k_i$ of the matrix of the dynamical system:
\[(A_{M} -k_{i} E){\bf u}_{i} =0,\]
\[{\rm Det}(A_{M} -k_{i} E)=0,\]
where $E$ is the unit matrix.  Thanks to the block-diagonal structure of the matrix of the dynamical system $A_{M}$, its eigenvalues are determined by the characteristic equations in the corresponding two-dimensional planes, and the eigenvectors ${\bf u}_{k}^{(M)}$ corresponding to these eigenvalues lie pairwise in different phase planes:  $\left\{{\bf u}_{1}^{(M)}, {\bf u}_{2}^{(M)} \right\}\in \Sigma _{\Phi } $, $\left\{{\bf u}_{3}^{(M)} ,{\bf u}_{4}^{(M)} \right\}\in \Sigma _{\varphi } $.  This fact allows us to significantly simplify the qualitative analysis of the phase trajectories near a singular point $M_\alpha$ and to reduce it to a listing of combinations of characteristics of the dynamical system in the two-dimensional planes $\Sigma_\Phi$, $\Sigma_\varphi$.  If a singular point is found in a forbidden region, then, as a consequence of realness of the elements of the matrix of the dynamical system, each complex eigenvalue $k$ of it should correspond to its complex conjugate value $\bar{k}$, so that $k\bar{k}=|k|^2>0$. If the singular point is found in an inaccessible region of phase space, the latter condition cannot be satisfied. In this case, the conclusions of the qualitative theory are only conditionally applicable to the extent of the smallness of the radius of the forbidden region.  The specific behavior of a phase trajectory in these cases must be refined with the help of numerical integration of the dynamical equations.

According to the qualitative theory of differential equations (see [6, 7], for example), the radius-vector of the phase trajectory $r(\tau )=(x_{1} (\tau ),\ldots ,x_{n} (\tau ))$ in the vicinity of the singular point $M_{\alpha } \left(x_1^{(\alpha)} ,\ldots ,x_{n}^{(\alpha )} \right)$ is described by the equation
\begin{equation} \label{_15_}
r(t)\simeq r^{(\alpha )} +\Re \left(\sum _{j=1}^{n}C_{j} {\bf u}_{j}^{(\alpha )} e^{ik_{j}^{(\alpha )} \tau } \right),
\end{equation}
where $C_{j}$ are arbitrary constants, determined by the initial conditions, $k_{j}^{(\alpha )}$  are the eigenvalues of the matrix of the dynamical system $A(M_\alpha)$, and ${\bf u}_j^{(\alpha)}$ are the eigenvectors of this matrix, corresponding to the eigenvalues $k_{j}^{(\alpha )}$.  In these cases, when the singular point is inaccessible, an estimative formula, nevertheless, is a sufficiently good approximation of the phase trajectory.  We will base ourselves on this estimate in cases when results of a standard qualitative theory of differential equations is lacking, which would be suitable for real matrices of the dynamical system.

Let us briefly enumerate the results of a qualitative analysis of dynamical system \eqref{_1_}. First of all, the calculations show that all of the singular points of the dynamical system divide into four groups, the character of the points inside each group being the same.

Only one singular point enters into the first group:
$$M_0(0,0,0,0);\quad \alpha\in \mathbb{R}, \quad \beta\in \mathbb{R}.$$

According to the qualitative theory of differential equations, the singular point $M_0$  can have the following character, depending on the parameters of the field model (Tables 1 and 2)\footnote{\ For\ details,\ see\ [3].\ \ \ \ \ }.
\begin{center}
\noindent TABLE 1. Character of the Singular Point $M_{0} $ in the \textbf{$\Sigma _{\Phi } $} Plane\\[12pt]
\begin{tabular}{p{1.2in}|p{0.4in}|p{1.2in}|p{1.2in}} \hline
$\lambda _{m} $ & $e$ & $k$ & Type \\ \hline
$\lambda _{m} =0$ & +1 & $\pm i$ & Center \\
 & -1 & $\pm 1$ & Saddle \\ \hline
$\lambda _{m} >0$ & -1 & $k_{1} >0;k_{2} <0$ & Saddle \\
$\lambda _{m} >4/3$ & +1 & $k_{1} <0;k_{2} <0$ & Attractive Node  \\
$0<\lambda _{m} <4/3$ & +1 & $\Re (k)<0$ & Attractive Focus    \\ \hline
$\lambda _{m} <0$ & +1 & $\Re (k_{1} )=\Re (k_{2} )=0$ & Inaccessible Center     \\
$-4/3<\lambda _{m} <0$ & -1 & $\Re (k_{1} )=-\Re (k_{2} )$ & Inaccessible Saddle  \\
$\lambda _{m} <-4/3$ & -1 & $\Re (k_{1} )=\Re (k_{2} )=0$ & Inaccessible Center   \\ \hline
\end{tabular}
\end{center}
\begin{center}
TABLE 2. Character of the Singular Point \textbf{$M_{0} $} in the $\Sigma _{\varphi } $ Plane\\[12pt]

\begin{tabular}{p{1.2in}|p{0.4in}|p{1.2in}|p{1.2in}} \hline
$\lambda _{m} $ & $\varepsilon $ & $k$ & Type \\ \hline
$\lambda _{m} =0$ & +1 & $\pm \mu $ & Saddle Point \\
 & -1 & $\pm i\mu $ & Center \\ \hline
$\lambda _{m} >0$ & +1 & $k_{3} >0;k_{4} <0$ & Saddle Point \\
$\lambda _{m} >4/3\mu ^{2} $ & -1 & $k_{3} <0;k_{4} <0$ & Attractive Node   \\
$0<\lambda _{m} <4/3\mu ^{2} $ & -1 & $\Re (k)<0$ & Attractive Focus   \\ \hline
$\lambda _{m} <0$ & -1 & $\Re (k_{3} )=\Re (k_{4} )=0$ & Inaccessible Center  \\
$-4/3\mu ^{2} <\lambda _{m} <0$ & +1 & $\Re (k_{3} )=-\Re (k_{4} )$ & Inaccesible Saddle  \\
$\lambda _{m} <-4/3\mu ^{2} $ & +1 & $\Re (k_{3} )=\Re (k_{4} )=0$ & Inaccessible Center  \\ \hline
\end{tabular}
\end{center}
The second group includes two symmetric singular points $M_{01},M_{02}$ under the condition $\varepsilon\beta_m>0$ $\forall\alpha_m$ $M_{01},M_{02}$:
\[M_{01} \left(0,0,\frac{\mu }{\sqrt{\varepsilon \beta _{m} } } ,0\right),\quad M_{02} \left(0,0,-\frac{\mu }{\sqrt{\varepsilon \beta _{m} } } ,0\right).\]
Tables 3 and 4 list possible types of these points.
\pagebreak
\begin{center}
\noindent TABLE 3. Character of the Singular Points \textbf{$M_{01} ,M_{02} $} in the \textbf{$\Sigma _{\Phi } $} Plane\\[12pt]
\begin{tabular}{p{1.2in}|p{0.4in}|p{1.2in}|p{1.2in}} \hline
$\sigma _{1}^{2} $ & $e$ & $k$ & Type \\ \hline
$\sigma _{1}^{2} =0$ & +1 & $\pm i$ & Center \\
 & -1 & $\pm 1$ & Saddle Point \\ \hline
$\sigma _{1}^{2} >0$ & -1 & $k_{1} >0;k_{2} <0$ & Saddle Point \\
$\sigma _{1}^{2} >1$ & +1 & $k_{1} <0;k_{2} <0$ & Attractive Node  \\
$0<\sigma _{1}^{2} <1$ & +1 & $\Re (k)<0 $ & Attractive Focus  \\ \hline
$\sigma _{1}^{2} <0$ & +1 & $\Re (k_{1} )=\Re (k_{2} )=0$ & Inaccessible Center  \\
$-1<\sigma _{1}^{2} <0$ & -1 & $\Re (k_{1} )=-\Re (k_{2} )$ & Inaccessible Saddle \\
$\sigma _{1}^{2} <-1$ & -1 & $\Re (k_{1} )=\Re (k_{2} )=0$ & Inaccessible Center  \\ \hline
\end{tabular}
\end{center}

\begin{center}
TABLE 4. Character of the Singular Points \textbf{$M_{01} ,M_{02} $} in the $\Sigma _{\varphi } $ Plane\\[12pt]
\begin{tabular}{p{1.2in}|p{0.4in}|p{1.2in}|p{1.2in}} \hline
$\sigma _{1}^{2} $ & $\varepsilon $ & $k$ & Type \\ \hline
$\sigma _{1}^{2} =0$ & +1 & $\pm i\sqrt{2} \mu $ & Center \\
 & -1 & $\pm \sqrt{2} \mu $ & Saddle Point \\ \hline
$\sigma _{1}^{2} >0$ & -1 & $k_{3} >0;k_{4} <0$ & Saddle Point \\
$\sigma _{1}^{2} >2\mu ^{2} $ & +1 & $k_{3} <0;k_{4} <0$ & Attractive Node  \\
$0<\sigma _{1}^{2} <2\mu ^{2} $ & +1 & $\Re (k)<0$ & Attractive Focus   \\ \hline
$\sigma _{1}^{2} <0$ & +1 & $\Re (k_{3} )=\Re (k_{4} )=0$ & Inaccessible Center   \\
$-2\mu ^{2} <\sigma _{1}^{2} <0$ & -1 & $\Re (k_{3} )=-\Re (k_{4} )$ & Inaccessible Saddle  \\
$\sigma _{1}^{2} <-2\mu ^{2} $ & -1 & $\Re (k_{3} )=\Re (k_{4} )=0$ & Inaccessible Center  \\ \hline
\end{tabular}
\end{center}

The third group of singular points includes the points $M_{10} $ and $M_{20} $ under the condition:\\ 
$M_{10} \left({\displaystyle\frac{1}{\sqrt{e\alpha _{m} }}} ,0,0,0\right)$, $M_{20} \left(-{\displaystyle\frac{1}{\sqrt{e\alpha _{m}  }}} ,0,0,0\right)$. 
Tables 5 and 6 list possible types of these points.

\begin{center}
\noindent TABLE 5. Character of the Singular Points $M_{10} ,M_{20} $ in the $\Sigma _{\Phi } $ Plane\\[12pt]
\begin{tabular}{p{1.2in}|p{0.4in}|p{1.2in}|p{1.2in}} \hline
$\sigma _{2} {}^{2} $ & $e$ & $k$ & Type \\ \hline
$\sigma _{2} {}^{2} =0$ & +1 & $\pm \sqrt{2} $ & Saddle Point \\
 & -1 & $\pm i\sqrt{2} $ & Center \\ \hline
$\sigma _{2} {}^{2} >0$ & +1 & $k_{1} >0;k_{2} <0$ & Saddle Point \\
$\sigma _{2} {}^{2} >2$ & -1 & $k_{1} <0;k_{2} <0$ & Attractive Node   \\
$0<\sigma _{2} {}^{2} <2$ & -1 & $\Re (k)<0$ & Attractive Focus   \\ \hline
$\sigma _{2} {}^{2} <0$ & -1 & $\Re (k_{1} )=\Re (k_{2} )=0$ & Inaccessible Center   \\
$-2<\sigma _{2} {}^{2} <0$ & +1 & $\Re (k_{1} )=-\Re (k_{2} )$ & Inaccessible Saddle \\
$\sigma _{2} {}^{2} <-2$ & +1 & $\Re (k_{1} )=\Re (k_{2} )=0$ & Inaccessible Center  \\ \hline
\end{tabular}
\end{center}

\begin{center}
TABLE 6. Character of the Singular Points $M_{10} ,M_{20} $ in the $\Sigma _{\varphi } $ Plane \\[12pt]
\begin{tabular}{p{1.2in}|p{0.4in}|p{1.2in}|p{1.2in}} \hline
$\sigma _{2} {}^{2} $ & $\varepsilon $ & $k$ & Type \\ \hline
$\sigma _{2} {}^{2} =0$ & +1 & $\pm \mu $ & Saddle Point \\
 & -1 & $\pm i\mu $ & Center \\ \hline
$\sigma _{2} {}^{2} >0$ & +1 & $k_{3} >0;k_{4} <0$ & Saddle Point \\
$\sigma _{2} {}^{2} >\mu ^{2} $ & -1 & $k_{3} <0;k_{4} <0$ & Attractive Nodoe  \\
$0<\sigma _{2} {}^{2} <\mu ^{2} $ & -1 & $\Re (k)<0$ & Attractive Focus  \\ \hline
$\sigma _{2} {}^{2} <0$ & -1 & $\Re (k_{3} )=\Re (k_{4} )=0$ & Inaccessible Center  \\
$-\mu ^{2} <\sigma _{2} {}^{2} <0$ & +1 & $\Re (k_{3} )=-\Re (k_{4} )$ & Inaccessible Saddle   \\
$\sigma _{2} {}^{2} <-\mu ^{2} $ & +1 & $\Re (k_{3} )=\Re (k_{4} )=0$ & Inaccessible Center   \\ \hline
\end{tabular}
\end{center}

Finally, the fourth group of singular points includes the points $M_{11}$, $M_{22}$, $M_{12}$ and $M_{21}$ under the condition $e\alpha_m>0$ and $\varepsilon\beta_m>0$:
\[\begin{array}{c} {M_{11} \left({\displaystyle\frac{1}{\sqrt{e\alpha _{m} }} } ,0,{\displaystyle\frac{\mu }{\sqrt{\varepsilon \beta _{m} }} } ,0\right),\, \, \, \, \, M_{12} \left({\displaystyle\frac{1}{\sqrt{e\alpha _{m} } }} ,0,-{\displaystyle\frac{\mu }{\sqrt{\varepsilon \beta _{m} }} } ,0\right),{\rm \; }}\\[12pt] M_{21} \left(-{\displaystyle\frac{1}{\sqrt{e\alpha _{m} }} } ,0,{\displaystyle\frac{\mu }{\sqrt{\varepsilon \beta _{m} }} } ,0\right),  {M_{22} \left(-{\displaystyle\frac{1}{\sqrt{e\alpha _{m} }} } ,0,-{\displaystyle\frac{\mu }{\sqrt{\varepsilon \beta _{m} } }} ,0\right).} \end{array}\]
Tables 7 and 8 list the possible types of these points.

\begin{center}
\noindent TABLE 7. Character of the Singular Points $M_{11} ,M_{12} ,M_{21} ,M_{22} $ in the $\Sigma _{\Phi } $ Plane\\[12pt]
\begin{tabular}{p{1.2in}|p{0.4in}|p{1.2in}|p{1.2in}} \hline
$\sigma _{3} {}^{2} $ & $e$ & $k$ & Type \\ \hline
$\sigma _{3} {}^{2} =0$ & +1 & $\pm \sqrt{2} $ & Saddle Point \\
 & -1 & $\pm i\sqrt{2} $ & Center \\ \hline
$\sigma _{3} {}^{2} >0$ & +1 & $k_{1} >0;k_{2} <0$ & Saddle Point \\
$\sigma _{3} {}^{2} >2$ & -1 & $k_{1} <0;k_{2} <0$ & Attractive Node   \\
$0<\sigma _{3} {}^{2} <2$ & -1 & $\Re (k)<0$ & Attractive Focus   \\ \hline
$\sigma _{3} {}^{2} <0$ & -1 & $\Re (k_{1} )=\Re (k_{2} )=0$ & Inaccessible Center   \\
$-2<\sigma _{3} {}^{2} <0$ & +1 & $\Re (k_{1} )=-\Re (k_{2} )$ & Inaccessible Saddle  \\
$\sigma _{3} {}^{2} <-2$ & +1 & $\Re (k_{1} )=\Re (k_{2} )=0$ & Inaccessible Center   \\ \hline
\end{tabular}
\end{center}

\begin{center}
TABLE 8. Character of the Singular Points $M_{11} ,M_{12} ,M_{21} ,M_{22} $ in the $\Sigma _{\varphi } $ Plane\\[12pt]
\begin{tabular}{p{1.2in}|p{0.4in}|p{1.2in}|p{1.2in}} \hline
$\sigma _{3} {}^{2} $ & $\varepsilon $ & $k$ & Type \\ \hline
$\sigma _{3} {}^{2} =0$ & +1 & $\pm i\sqrt{2} \mu $ & Center \\
 & -1 & $\pm \sqrt{2} \mu $ & Saddle Point \\ \hline
$\sigma _{3} {}^{2} >0$ & -1 & $k_{3} >0;k_{4} <0$ & Saddle Point \\
$\sigma _{3} {}^{2} >2\mu ^{2} $ & +1 & $k_{3} <0;k_{4} <0$ & Attractive Node  \\
$0<\sigma _{3} {}^{2} <2\mu ^{2} $ & +1 & $\Re (k)<0$ & Attractive Focus  \\ \hline
$\sigma _{3} {}^{2} <0$ & +1 & $\Re (k_{3} )=\Re (k_{4} )=0$ & Inaccessible Center   \\
$-2\mu ^{2} <\sigma _{3} {}^{2} <0$ & -1 & $\Re (k_{3} )=-\Re (k_{4} )$ & Inaccessible Saddle \\
$\sigma _{3} {}^{2} <-2\mu ^{2} $ & -1 & $\Re (k_{3} )=\Re (k_{4} )=0$ & Inaccessible Center  \\ \hline
\end{tabular}
\end{center}

\section{ Maps of singular points of the dynamical system of an asymmetric scalar doublet      }

To summarize, we have enumerated all the characteristics of the singular points in the two-dimensional planes $\Sigma_\Phi$ and $\Sigma_\varphi$.  To obtain a general, four-dimensional characteristic of each singular point, it is necessary to multiply their characteristics in the corresponding planes.  But, in this case it is necessary to confirm that the corresponding conditions on the parameters are noncontradictory.  For example, let us consider the noncontradictory cases $\sigma_3^2>2$, $e=-1$ from Table 7 and $\sigma_3^2>0$, $\varepsilon=-1$ from Table 8.  Together, they correspond to a singular point which is an attractive node in the $\Sigma_\Phi$ plane and a saddle point in the $\Sigma_\varphi$ plane.   But, for the existence of such a singular point, it is necessary that $\alpha_m<0,\beta_m<0$.  Now we must consider to what situation the condition $\sigma_3^2<2$ and the condition stated by inequality \eqref{_12_}  lead, in order to clarify in which region the singular point is found.  \textit{Maps of the singular points of the dynamical system} help us make sense of such a complicated picture.

Figure 1 presents a map of the singular points of the dynamical system with the parameters
\begin{equation}\label{_16_}
{\bf P}=[1,1,1,1,1,0].
\end{equation}

\noindent Such maps reflect the character of the singular points of the dynamical system.  Here and in what follows, the dark-gray color depicts centers, the light-gray color depicts attractive nodes, the gray color depicts attractive foci, and the white color depicts saddle points.  The left half of the circle corresponds to the character of the point in the $\Sigma_\Phi$ plane, and the right half corresponds to the character of the  point in the $\Sigma_\varphi$  plane.  Black color of the boundary of the circle corresponds to an inaccessible point, and gray color, to an accessible point\footnote{\ For\ the\ corresponding\ maps\ in\ color,\ see\ [4].\ In\ the\ black-and-white\ format\ of\ the\ figure,\ it\ is\ not\ possible\ to\ unambiguously\ reflect\ the\ character\ of\ the\ points.\ In\ our\ figure,\ blue\ corresponds\ to\ the\ dark-gray\ shading,\ and\ red\ corresponds\ to\ the\ light-gray\ shading.\ }.

 \begin{figure}[h!]
 \centerline{\includegraphics[width=8cm]{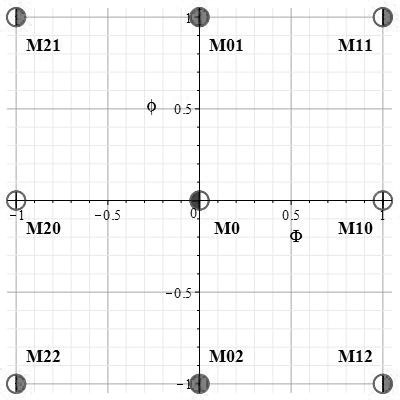}\label{Fig2}} \caption{ Map of singular points for the parameters of the model specified by Eq. (16).
 }
\end{figure}

This work was performed within the scope of the Russian Government Program of Competitive Growth of Kazan Federal University.

\end{document}